# What Malaysian Software Students Think about Testing?

Luiz Fernando Capretz
Electrical & Computer Engineering
Western University
London, Ontario, Canada
lcapretz@uwo.ca

Shuib Basri    Maythem Adili
Computer & Information Sciences
Universiti Teknologi Petronas
Seri Iskander, Perak, Malaysia
{shuib_basri, maythem.aladili}@utp.edu.my

Aamir Amin
Information Systems Department
University Tunku A. Rahman
Kampar, Perak, Malaysia
aamir@utar.edu.my

## ABSTRACT

Software testing is one of the crucial supporting processes of software life cycle. Unfortunately for the software industry, the role is stigmatized, partly due to misperception and partly due to treatment of the role in the software industry. The present study aims to analyse this situation to explore what inhibit an individual from taking up a software testing career. In order to investigate this issue, we surveyed 82 senior students taking a degree in information technology, information and communication technology, and computer science at two Malaysian universities. The subjects were asked the PROs and CONs of taking up a career in software testing and what were the chances that they would do so. The study identified 7 main PROs and 9 main CONSs for starting a testing career, and indicated that the role of software tester has been perceived as a social role, with more soft skills connotations than technical implications. The results also show that Malaysian students have a more positive attitude towards software testing than their counterparts where similar investigations have been carried out.

## CCS CONCEPTS

• Software engineering • Software testing • Human factors in software engineering

## KEYWORDS

Human factors in software engineering, Software engineering, Software testing, Software careers, Cross-cultural studies

## 1   Introduction

It has been pointed out that human and social aspects play a significant role in software testing practices. Attention to human factors in software testing has been investigated in professional settings [1] [2] and academic environments [3] [4]. We started asking whether students would consider to being a software tester, then their opinions about a software testing career.

## 2   Methodology

The survey asked four questions. The first two questions were open ended questions: 1) What are three PROs (in order of importance) of pursuing a career in software testing; and 2) What are three CONs (also prioritized) of pursuing a career in software testing. The third question asked participants to indicate their intentions of pursuing a career in software testing. They were given the option to answer with either "certainly not," "no," "maybe," "yes," and "certainly yes." Participants were also invited to share the reasons behind their responses.

Our study analyzed the reasons for not choosing testing careers by Malaysian students. These are senior students taking a degree in Information Communication and Technology, Information Technology, and Computer Science at Universiti Teknologi PETRONAS (UTP) and University Tunku Abdul Rahman (UTAR), both in Malaysia. Eighty two (82) students voluntarily participated in the survey; 35 of participants were males, 40 were females, 7 preferred not to disclose.

## 3   Results

Table I shows the responses to the actual chances of respondents taking up a testing career according to their personal preferences.

TABLE I. CHANCES TO TAKE UP TESTING CAREERS

| Responses (82) | Numbers | Percentage |
|---|---|---|
| Certainly Not | 1 | 1% |
| No | 7 | 7% |
| May be | 52 | 52% |
| Yes | 34 | 34% |
| Certainly Yes | 6 | 6% |
| **TOTAL** | **82** | **100%** |

Similar statements were combined during the refining of data. We found 7 main PROs and 9 main CONs in total; these statements are listed in Table II and Table III below. The most important reasons considered as PROs for taking up a testing career among the surveyed individuals are presented in Table II, along with their frequencies.

In contrast, when asked about the CONs for taking up a testing career, respondents gave most importance to the following reasons: (a) it is a tedious job for 39% of respondents, (b) it is a very difficult and complex task for 33%; and (c) 34% of the respondents pointed out the perception that other team members may become upset facing tester´s findings, i.e. software failures. Lastly, 6% of the subjects noted that in the labor market the role of tester is a role for which wages are lower than the average wages for other roles; that view is reinforced by the perception that software testers are treated as second-class citizens within a software project (7%).

TABLE II. PERCENTAGES OF SALIENT PROs

| PROs | Percentages |
|---|---|
| Learning Opportunities | 53% |
| Important Job | 37% |
| Easy Job | 28% |
| Thinking job | 20% |
| More jobs | 37% |
| Monetary benefits | 29% |
| Fun to break things | 7% |

The CONs responses are analyzed and presented in Table III. Since we excluded CONs that were less than 5% of the total CONs, the total in each column may not be 100%. Moreover, respondents could list more than one reason.

TABLE III. PERCENTAGES OF SALIENT CONs

| CONs | Percentages |
|---|---|
| Second-class citizen | 7% |
| Career development | 29% |
| Complexity | 33% |
| Tedious | 39% |
| Missing development | 18% |
| Less monetary benefits | 6% |
| Finding others' mistakes | 34% |
| Stressful job | 15% |
| No interest | 17% |

## 4 Discussions and Implications

The study certainly offers useful insights and helps educators to come up with an action plan to change the outlook towards testers in their programs, and put the software testing profession under a new light. That could increase the number of software engineers deciding on testing as a career of their choice, could increase the quality of software testing, and improve the overall productivity, and turnaround time of software development activity.

The study has many implications for colleges, especially for information sciences, information technology, computer science and software engineering programs. Since testing courses can improve the perception of testing careers, universities can introduce them in their curricula. They can regularly review the curricula by consulting their alumni and to ongoing research. Since testing offers additional jobs, the course can help colleges improve placement prospects of their students.

The testing curriculum needs to reflect the understanding that testers need to provide correct information to various stakeholders, and appreciate that testing is 'applied epistemology' grounded in 'social psychology'. Course instructors must dispel beliefs such as, testing is just mechanically running tests and comparing outputs with expected results. Instead, they should explain the importance of testing and the philosophy behind it and impress upon the students that any testing assignment and design of test cases can be very creative. They should develop testers who can understand different domains and the needs of users in those domains, to understand the developer mindset and anticipate mistakes that developers may be making as an individual and as a team, to test creatively and efficiently under the given constraints, and report the findings tactfully to all stakeholders.

The general empirical findings on the advantages as perceived by students varies considerably. While learning opportunities and job opportunities and importance are the common advantages, their percentages vary widely from 7% to 53% among the main PROs. Additionally, the PROs and CONs of testing careers, as perceived by students, can help test managers and team leaders scale up the challenge of recruiting test professionals. Understanding the common as well as country-specific advantages and drawbacks may help managers dealing with global teams. As emphasized before, software testing is a human activity [5] [6] and testers, who willingly take up testing careers [7] [8] [9], can influence the quality of the final product.

With 34% of students indicating that they would like to be a software tester, and 6% would certainly take up a career as software tester, it is evident that a software testing profession among students is not as unpopular as it is in other countries [10]; whereas only 7% of the students would not like to be software tester. These results do not concur with prior studies [11] [12], which point to the tester role as one of the least popular roles among others, such as project manager, analyst, designer, programmer, and maintainer.